\documentclass[aps,prl,groupedaddress,showpacs,preprint]{revtex4}
\usepackage{graphicx}
\usepackage[dvips]{epsfig}
\usepackage{dcolumn}
\usepackage{subfigure}%
\usepackage{bm}
\usepackage{amssymb}
\usepackage{amsmath}

\begin{document}
\title{Dynamic Scaling of Non-Euclidean Interfaces}

\author{Carlos Escudero}

\affiliation{Mathematical Institute, University of Oxford, 24-29 St Giles', Oxford OX1 3LB, United Kingdom}

\begin{abstract}
The dynamic scaling of curved interfaces presents features that are strikingly different from those of the planar ones. Spherical surfaces above one dimension are flat because the noise is irrelevant in such cases. Kinetic roughening is thus a one-dimensional phenomenon characterized by a marginal logarithmic amplitude of the fluctuations. Models characterized by a planar dynamical exponent $z>1$, which include the most common stochastic growth equations, suffer a loss of correlation along the interface, and their dynamics reduce to that of the radial random deposition model in the long time limit. The consequences in several applications are discussed, and we conclude that it is necessary to reexamine some experimental results in which standard scaling analysis was applied.
\end{abstract}

\maketitle

In Nature, one finds a large number of phenomena driven by growing interfaces. To cite a few, we can recall wetting fronts, flames, bacterial growth, fluid flow in porous media, thin-film deposition.... All these, apparently independent, processes have been studied under the same theoretical framework: scaling analysis~\cite{barabasi}. Scaling analysis characterizes growing surfaces with sets of critical exponents, encoding information about the morphology and dynamics of the interfaces. Different phenomena sharing the same critical exponents are said to belong to the same universality class. Thus, classifying in terms of universality classes allows us to identify how growth dynamics with a very diverse origin can have the main physical mechanism in common. Despite its successes, there are some limitations associated with scaling analysis so far. For instance, the interfaces are supposed to be describable from a Euclidean reference frame, and they do not alter their size during the evolution. While these assumptions are reasonable in many cases, on can find as many where they do not hold. Biology, for instance, is remarkable for its strong tendency towards the approximate spherical symmetry: bacterial colonies~\cite{vicsek}, fungi~\cite{matsuura}, plant calli~\cite{galeano}, and tumors~\cite{bru1} develop rough interfaces that escape the hypothesis of traditional scaling analysis. Not only biological systems, but also physical settings such as fluid flow in porous media~\cite{barabasi}, grain-grain displacement in Hele-Shaw cells~\cite{ferreira}, adatom and vacancy islands on surfaces~\cite{einstein1}, and atomic ledges bordering crystalline facets~\cite{ferrari,einstein2} present interfaces that either become larger as time evolves or have a non-Euclidean geometry. For this reason, this problem has been considered as one of the main open questions of scaling analysis~\cite{cuerno}.

Traditionally, systems like flames or wetting fronts have been studied in strip geometry in well controlled experiments~\cite{barabasi} although they appear more commonly as propagating in all directions from a small, pointlike region. Even bacterial colonies have been grown in this fashion~\cite{vicsek}, without knowing the relation between the scaling behavior in this geometry and the radial one that appears naturally. Tumors have been an exception to this rule~\cite{bru1} since they were grown from a seed and they acquired a radial interface during the evolution. However, standard scaling analysis was applied to this case as well, a fact that is being increasingly questioned in the literature~\cite{galeano,ferreira,drasdo,buceta,brutovsky}. These last two systems are particularly important in applications, due to the possibility of understanding and even controlling bacterial infections and malignant tumor proliferation. Indeed, a strategy to stop tumor growth, based on the planar scaling analysis in~\cite{bru1}, was proposed in~\cite{bru2}. Following this reference, an enhancement of the immune response would be able to slow down tumor growth, and it would eventually lead to the pinning of the tumor interface. However, these results enter in contradiction with others present in the medical literature. In~\cite{larco}, it is proposed that the infiltration of a tumor by neutrophils, the most common immune cell, can promote tumor progression and the appearance of metastases. Furthermore, in this article, a possible therapy is proposed, consisting in inhibiting the chemical {\it interleukin-8}, responsible for the attraction of the neutrophils to the tumor; this is exactly the opposite strategy to that proposed in~\cite{bru2}.

The theoretical study of non-equilibrium radial growth dates back to Eden~\cite{eden}, but the use of stochastic growth equations appeared only more recently. To our knowledge, the first work in this direction is~\cite{maritan}, where the authors proposed the Kardar-Parisi-Zhang (KPZ) equation in reparametrization invariant form. After that, several works appeared studying the radial (2D) KPZ equation~\cite{livi,batchelor2}, and the radial (2D) and spherical (3D) Mullins-Herring (MH) equation~\cite{escudero1,escudero2}. The equation of growth of a general Riemannian surface reads
\begin{equation}
\partial_t \vec{r}({\bf s},t)=\hat{n}({\bf s},t)\Gamma[\vec{r}({\bf s},t)]+\vec{\Phi}({\bf s},t),
\end{equation}
where the $d+1$ dimensional surface vector $\vec{r}({\bf s},t)=\{r_\alpha({\bf s},t)\}_{\alpha=1}^{d+1}$ runs over the surface as ${\bf s}=\{s^i\}_{i=1}^d$ varies in a parameter space. In this equation $\hat{n}$ stands for the unitary vector normal to the surface at $\vec{r}$, $\Gamma$ contains a deterministic growth mechanism that causes growth along the normal $\hat{n}$ to the surface, and $\vec{\Phi}$ is a random force acting on the surface. The stochastic term comes from the force in the same way as its deterministic counterpart $\vec{\Phi}=\hat{n} \eta$. The noise $\eta$ is assumed to be a Gaussian variable with zero mean and correlation given by
\begin{equation}
\label{correlation}
\left< \eta({\bf s},t) \eta({\bf s}',t') \right> = n_\alpha({\bf s},t) n_\beta({\bf s},t) \epsilon \delta^{\alpha \beta} g^{-1/2}\delta({\bf s}-{\bf s}')\delta(t-t'),
\end{equation}
where the Einstein summation convention has been adopted. We will now focus on the radial geometry
$\vec{r}=\left(r(\theta,t)\cos(\theta),r(\theta,t)\sin(\theta)\right)$, and will start analyzing the simplest process: random deposition. In a Euclidean geometry, the corresponding stochastic growth equation is $\partial_t h = F + \eta({\bf x},t)$, where $F$ is the constant deposition rate, and $\eta$ is a zero mean Gaussian noise with correlation given by
$\left< \eta({\bf x},t)\eta({\bf x}',t') \right>= \epsilon \delta({\bf x}-{\bf x}')\delta(t-t')$. In this case, the only well defined exponent is the growth one $\beta = 1/2$, which is superuniversal, i. e., independent of the spatial dimension. This value can be straightforwardly obtained from the properties of white noise: after time integration, we obtain a temporal Wiener process in every spatial point; as its variance increases linearly in time, we obtain the desired result averaging over uncorrelated spatial points, this is, the interface width evolves as
\begin{equation}
W({\bf L},t)= \left< \overline{\left[h({\bf x},t)-{\bar h}\right]^2} \right>^{1/2} \sim t^{1/2},
\end{equation}
for all times, where the overbar denotes spatial average over the fixed system size ${\bf L}=(L_1,\cdots,L_d)$, and the brackets the average over different realizations. In the case of a one-dimensional radial geometry, the growth equation reads~\cite{livi,escudero1,escudero2}
\begin{equation}
\partial_t r = F + r^{-1/2}\eta(\theta,t),
\end{equation}
after the usual small gradient expansion, that assumes $|\partial_\theta r| \ll r$~\cite{livi}. The noise correlation is given by $\left< \eta(\theta,t)\eta(\theta',t') \right>= \epsilon \delta(\theta-\theta')\delta(t-t')$, but due to its multiplicative nature, we need to add an interpretation, which is unambiguously It\^o. This is because the multiplicative term appears as a consequence of the reparametrization invariance of the correlation, Eq.(\ref{correlation}), while the process has zero mean. The interface width can be defined in this case as~\cite{ferreira,brutovsky}
\begin{equation}
W_r(t)=\left< \overline{\left[r(\theta,t)-{\bar r}\right]^2} \right>^{1/2},
\end{equation}
expression that will be evaluated with the help of a small noise expansion~\cite{escudero2}: $r(\theta,t)=R(t)+\sqrt{\epsilon}\rho(\theta,t)$, where the noise amplitude $\epsilon$ is assumed to be small enough. We get $R(t)=Ft$, for $t > t_0$, and
\begin{equation}
\label{rd}
\partial_t \rho = (Ft)^{-1/2}\xi(\theta,t),
\end{equation}
where $\eta=\sqrt{\epsilon}\xi$, that is actually an ordinary stochastic differential equation. The corresponding Fokker-Planck equation for its probability distribution function is
\begin{equation}
\partial_t P(\rho,t) = (4 \pi Ft)^{-1}\partial_{\rho}^2 P(\rho,t),
\end{equation}
that, by changing the temporal variable $\tau= \mathrm{ln}(t)$, is reduced to the diffusion equation. So we see that in this case, the variance is proportional to the logarithm of time, and averaging over uncorrelated angular points, we get $W_r(t) \sim [\mathrm{ln}(t/t_0)]^{1/2}$ for all times, while for short times $t \gtrsim t_0$, $W_r(t) \sim t^{1/2}$, a behavior reminiscent to that of the planar case. However, the long time behavior shows a marked weakening of the fluctuations, and, as we will show, the logarithmic dependence is the footprint of critical dimensionality. The two-dimensional random deposition equation is~\cite{escudero2}
\begin{equation}
\partial_t r = F + \left( r \left| \sin(\theta) \right|^{1/2} \right)^{-1}\eta(\theta,\phi,t),
\end{equation}
where the zero mean Gaussian noise has the correlation $\left< \eta(\theta,\phi,t)\eta(\theta',\phi',t') \right>= \epsilon \delta(\theta-\theta')\delta(\phi-\phi')\delta(t-t')$. The polar coordinate system $\vec{r}=\left( r(\theta,\phi)\sin(\theta)\cos(\phi),r(\theta,\phi)\sin(\theta)\sin(\phi),r(\theta,\phi)\cos(\theta) \right)$ parameterizes the equator of the sphere for $\theta=\pi/2$. In this case, the equation for the small perturbation $\sqrt{\epsilon}\rho(\theta=\pi/2,\phi,t)=r(\theta=\pi/2,\phi,t)-Ft$ is $\partial_t \rho = (Ft)^{-1}\xi(\pi/2,\phi,t)$, which Fokker-Planck equation
\begin{equation}
\partial_t P(\rho,t) = (4 \pi)^{-1}(Ft)^{-2}\partial_{\rho}^2 P(\rho,t),
\end{equation}
is reduced to the diffusion equation by means of changing the temporal variable $\tau = -t^{-1}$. Averaging over the uncorrelated angular points on the equator, we get $W_r(t) \sim t_0^{-1} - t^{-1}$ for all times, which reveals a fact that is completely new to this geometry: the width saturates in the case of the two-dimensional random deposition process, and the surface becomes flat after a transient when some residual roughness develops. This result can be easily generalized to any spatial dimension $d$: the Langevin equation for the small stochastic perturbation for $d \ge 2$ is
\begin{equation}
\partial_t \rho = (Ft)^{-d/2}\xi(\phi,t),
\end{equation}
when restricted to the equator, parameterized by the angle $\phi$. The Fokker-Planck equation,
\begin{equation}
\partial_t P(\rho,t) = (4 \pi)^{-1} (Ft)^{-d}\partial_{\rho}^2 P(\rho,t),
\end{equation}
is related to the diffusion equation by the change of variable $\tau = -t^{1-d}$. So we obtain the evolution of the interface width $W_r(t) \sim t_0^{1-d} - t^{1-d}$ for $d \ge 2$, revealing that the surface becomes flat after a transient that is shorter for higher dimensions. The consequence of this fact is very important: $d=1$ is the critical dimension for the radial random deposition process. Above this dimension, the interface becomes flat, and the one-dimensional setting shows critical logarithmic fluctuations. Another interesting feature is that the scaling properties of the equator, a one-dimensional cut across the $d$-dimensional sphere, are those of the higher dimensional manifold in which it is embedded. The differences with the planar case come from the fact that radial interfaces grow in all directions. So, while we could picture the random deposition process as the growth of uncorrelated columns in the planar case, the sources of noise are distributed in a growing area in radial processes. This rends the fluctuations weaker for higher spatial dimensions.

Let us now focus on the radial Edwards-Wilkinson (EW) equation~\cite{livi,batchelor2}
\begin{equation}
\partial_t r = D r^{-2} \partial_\theta^2 r - D r^{-1} + F + r^{-1/2}\eta(\theta,t).
\end{equation}
The small noise expansion $r=R+\sqrt{\epsilon}\rho$ provides us with an equation for the deterministic contribution $\dot{R} = F - D R^{-1}$, that can be solved to yield
\begin{equation}
\label{exact}
R(t)=DF^{-1}\left\{ 1 + \mathcal{W}\left[ \left( F^2 t_0 D^{-1}-1 \right) \exp \left( F^2 t D^{-1} -1 \right) \right] \right\},
\end{equation}
where $\mathcal{W}$ is the Lambert omega function, and we have assumed $R(t_0)=Ft_0$ as initial condition and $t \geq t_0$. An inspection of this formula reveals that the value $F^2 t_0 = D$ is a fixed point of the dynamics~\cite{livi}. Initial conditions below the threshold, $R(t_0) < D/F$, shrink till they collapse, while the supercritical behavior is characterized by an accelerated growth rate, that approaches the linear law $R(t) \sim Ft$ in the long time limit. Shrinking of subcritical solutions produces a collapse in finite time
\begin{equation}
T_c= D F^{-2} \mathrm{ln}\left[ D (D-F^2t_0)^{-1} \right]-t_0.
\end{equation}
In this limit $t \to T_c$, the solutions behave as $R \to 0$ and $\dot{R} \to - \infty$. This "bistability", expansion vs. collapse, can produce transitions from one of these states to the other when noise is present into the system, a possibility that was in explored in~\cite{livi} by means of the backward Kolmogorov equation and simulations of a lattice model. Indeed, the fluctuations can drive part of the interface from a general expanding behavior to the shrinking one, as observed in plant calli due to dehydration processes~\cite{galeano}. Despite the interest of the possible relation among calli shrinking and the radial EW equation, we will not consider here large fluctuations able to change the interface dynamics: rare event solutions of the full stochastic equation are not analytic in the small parameter, and are thus suppressed by the small noise expansion. We will assume henceforth a large enough initial radius, such that rare events able to shift the interface expansion are so infrequent that their appearance can be ignored during the evolution time of interest. The stochastic perturbation obeys the equation
\begin{equation}
\partial_t \rho = D(Ft)^{-2} \left( \partial_\theta^2 \rho + \rho \right) + (Ft)^{-1/2}\xi(\theta,t),
\end{equation}
in the long time limit. Expressing its mean value as a Fourier series $\left< \rho(\theta,t) \right>=\sum_n e^{int}\rho_n(t)$, we get for the Fourier modes
\begin{equation}
\rho_n(t)=\exp \left[ D F^{-2}(1-n^2)\left( t_0^{-1}-t^{-1} \right)\right]\rho_n(t_0),
\end{equation}
reflecting that the $n=0$ mode is unstable and the $n = \pm 1$ modes are marginal~\cite{batchelor2}. The effect of perturbing the zeroth mode can be estimated by means of the exact solution Eq.(\ref{exact}), while the marginality of the other two is possibly related to the deviation from the circular shape of the growing interface of the lattice model~\cite{livi}. The rest of the modes is stable, but part of the perturbation persists even in the long time limit, a fact that can explain the higher complexity of the radial interface when compared to its planar counterpart~\cite{livi}. All of these features are new and due to the non-Euclidean geometry of the surface. We can also get an equation for the correlation
\begin{eqnarray}
C(\theta,\theta',t)&=&\sum_{n,m=-\infty}^\infty e^{i(n\theta+m\theta')}\left< \rho_n(t) \rho_m(t) \right>, \\
\frac{d}{dt}\left< \rho_n(t) \rho_m(t) \right>&=&\frac{D}{F^2t^2}(2-n^2-m^2)\left< \rho_n(t) \rho_m(t) \right>+\frac{\delta_{n,-m}}{2\pi Ft},
\end{eqnarray}
and solve it to obtain
\begin{eqnarray}
\nonumber
\left< \rho_n(t) \rho_m(t) \right>=\exp \left[ \frac{D}{F^2t}(m^2+n^2-2) \right]\left\{ \left< \rho_n(t_0) \rho_m(t_0) \right> \exp \left[ \frac{D}{F^2t_0}(2-m^2-n^2) \right] \right. \\
- \left. \frac{\delta_{n,-m}}{2 \pi F}\left( \mathrm{Ei} \left[ \frac{D}{F^2t}(2-m^2-n^2) \right]-\mathrm{Ei}\left[ \frac{D}{F^2t_0}(2-m^2-n^2) \right] \right) \right\},
\end{eqnarray}
where $\mathrm{Ei}$ denotes the exponential integral. Assuming an uncorrelated initial condition and using the asymptotic equivalence $\mathrm{Ei}(z) \sim \mathrm{ln} \left( |z| \right)$ when $z \sim 0$ yields
\begin{equation}
\left< \rho_n(t) \rho_m(t) \right> \sim \frac{\delta_{n,-m}}{2 \pi F} \mathrm{ln}(t), \qquad
C(\theta,\theta',t) \sim \frac{\mathrm{ln}(t)}{2 \pi F}\delta(\theta-\theta'),
\end{equation}
when $t \to \infty$. This shows that the interface becomes uncorrelated for long times, and it is exactly the same result that we would obtain from Eq.(\ref{rd}). So, except for the bistability expansion-collapse, the radial EW dynamics reduces to random deposition for long times, a fact that can be easily generalized to an arbitrary dimension. Not only EW, but also the simpler radial MH dynamics~\cite{escudero1}, that shows a monotone linear growth behavior, reduces to random deposition in the long time limit, as can be seen upon applying the same argument to the correlation function~\cite{escudero2}. The physical origin of this asymptotic behavior is the diffusive (subdiffusive) propagation of the correlations of the EW (MH) dynamics, as given by the dynamical exponent $z=2$ ($z=4$) in the planar case, that, in presence of a linear increase of the system size, rends the interface uncorrelated.

The radial KPZ equation~\cite {livi, batchelor2}
\begin{equation}
\partial_t r = D r^{-2} \partial_\theta^2 r - D r^{-1} + F \left[ 1+2^{-1}r^{-2} \left( \partial_\theta r \right)^2 \right] + r^{-1/2}\eta(\theta,t),
\end{equation}
retains the first order nonlinearity in the derivatives of the radius accounting for lateral growth. The planar KPZ dynamical exponent $z=3/2$ shows sub-ballistic behavior, which suggests that, assuming the validity of the planar description locally for the radial process would yield the random deposition dynamics for long times. Let us show this explicitly. Considering just this new nonlinearity in the small perturbation approximation
\begin{equation}
\partial_t \rho =F 2^{-1}(Ft)^{-2} \left( \partial_\theta \rho \right)^2,
\end{equation}
shows that, upon applying an angular derivative to both sides of this equation and shifting the time variable $\tau=(Ft)^{-1}$, the gradient $\partial_\theta \rho(\theta,\tau)$ obeys Burgers equation. So perturbations are constant along characteristics in inverse time, implying in turn that they will be able to correlate, at most, a finite angle proportional to the size of the perturbation and to the inverse of the initial radius. This can be read from the exact implicit solution $\rho_\theta=\rho_\theta^{(0)}\left[\theta - \rho_\theta F^{-1} (t^{-1}-t_0^{-1}) \right]$, where $\rho_\theta^{(0)}$ denotes the initial condition. So, for large enough radius $t_0 \gg 1$, small perturbations will be able to correlate only part of the interface, that becomes shorter for larger initial times. The hydrodynamic limit will be again characterized by a radial random deposition behavior, after a perhaps long and complex transient when part of the planar KPZ exponents could be temporally recovered in some spatio-temporal scales~\cite{batchelor2}.

In conclusion, we have shown that the dynamic scaling of non-Euclidean interfaces presents strikingly new features. Growing surfaces above $d=1$ become flat after some transient because the noise is irrelevant in such cases. Only one-dimensional interfaces show nontrivial stochastic dynamics, but the noise is much weaker than that of the planar case: the random deposition interface width increases only as the square root of the logarithm of time. Our results also point to the fact that models showing sub-ballistic propagation of the correlations, corresponding to the value of the planar dynamical exponent $z>1$, reduce to the radial random deposition model in the long time limit, implying the loss of correlation of the interface. Taking into account that the most common stochastic growth equations fulfill $z>1$ in planar geometry, we expect that radial random deposition dynamics will be of broad importance to understand the physics of growing radial clusters. In~\cite{ferreira}, it is shown an interesting effect that appears in radial geometry: the value of the critical exponents might be different if we choose a static origin as reference or if we use the border center of mass. As noted by the authors, this is due to the nontrivial growth of the border center of mass fluctuations. According to our present analysis, the interface fluctuations trivialize in the case of higher dimensions, suggesting that this effect is genuinely one-dimensional. This fact can be related to the properties of adatom and vacancy islands on surfaces. These islands perform Brownian motion in equilibrium, with their diffusion constants being proportional to a negative power of their radius, and their interface dynamics being described by the EW and MH equations~\cite{einstein1}. Out of equilibrium, adatom clusters might grow by deposition of new atoms, while vacancy clusters grow due to evaporation processes. In this regime, it would be interesting to observe if the growing islands behave as described by the radial EW and MH equations, this is, showing an uncorrelated and logarithmically rough interface. Also, the behavior of a rounded crystal facet ledge might be of deep interest in order to understand KPZ dynamics. In equilibrium, the roughness exponent is $\alpha=1/3$ instead of the random walk one $\alpha=1/2$, as predicted theoretically~\cite{ferrari} and afterwards found experimentally~\cite{einstein2}. This reveals that a nonlinearity in the manner of KPZ is accounting for the equilibrium fluctuations. In the nonstationary situation, for growing ledges, we predict a dynamic behavior in the hydrodynamic limit that is totally analogous to that of an island. So this physical setting constitutes an excellent test that can clarify whether this hydrodynamic limit is unphysically large or experimentally accesible. The implications of our results in medicine can have deep consequences too. The planar scaling analysis of growing tumors~\cite{bru1} can lead to misleading results, and any therapeutic strategy based on it~\cite{bru2} should be taken with extreme caution. In light of our results, it seems prudent to reconsider other complex growth problems to assess the effects of geometry. As we have shown in the basic models considered here, the consequences can indeed be profound.

This work has been partially supported by the MEC (Spain) through Projects Nos. EX2005-0976 and FIS2005-01729.

\end{document}